\documentclass[aps,prd,twocolumn,amsmath,amssymb,amsfonts,nofootinbib,superscriptaddress,altaffilletter]{revtex4-1}
\usepackage{array}
\newcolumntype{P}[1]{>{\centCas Aering\arraybackslash}p{#1}}

\usepackage{graphicx}
\usepackage{cellspace}
\usepackage{amssymb}
\usepackage{amsmath}
\usepackage{longtable}
\usepackage{tabu}
\usepackage{color}
\usepackage{etoolbox}
\usepackage{supertabular}
\usepackage{dcolumn}
\usepackage{ wasysym }

\def\Tobs{T_{\textrm{\mbox{\tiny{obs}}}}}
\def\Tcoh{T_{\textrm{\mbox{\tiny{coh}}}}}
\def\Tref{T_{\textrm{\mbox{\tiny{ref}}}}}

\def\EatH{Einstein@Home}
\def\EatHs{Einstein@Home }

\def\sci#1#2{#1\times10^{#2}}

\newcommand{\avgSeg}[1]{\overline{#1}}			

\newcommand{\Freq}{f}
\newcommand{\fdot}{{\dot{\Freq}}}
\newcommand{\fddot}{\ddot{\Freq}}

\newcommand{\Gauss}{\mathrm{\MakeUppercase{G}}}
\newcommand{\Signal}{{\mathrm{\MakeUppercase{S}}}}
\newcommand{\Line}{{\mathrm{\MakeUppercase{L}}}}
\newcommand{\Transient}{{\mathrm{t\MakeUppercase{L}}}}

\newcommand{\NoisetL}{{\Gauss\Line\Transient}}

\providecommand{\sc}[1]{\widehat{#1}}
\renewcommand{\sc}[1]{\widehat{#1}}

\newcommand{\BSNtsc}{{\hat\beta}_{{\Signal/\NoisetL}}}	
\newcommand{\BSGLtLr}{{\hat\beta}_{{\Signal/\NoisetL r}}}

\newcommand{\F}{\mathcal{F}}		

\newcommand{\scF}{\sc{\F}}

\newcommand{\scFtho}{\scF_*^{(0)}}

\newcommand{\avF}{\avgSeg{\F}}

\newcommand{\Nseg}{{N_{\mathrm{seg}}}}

\newcommand{\posCasAa}{\ensuremath{6.1237704239609}}
\newcommand{\posVelaJra}{\ensuremath{2.3213891342490}}
\newcommand{\posGa}{\ensuremath{4.509370536464}} 
\newcommand{\posCasAd}{\ensuremath{{1.0264578036951}}} 
\newcommand{\posVelaJrd}{\ensuremath{-0.8080542824176}} 
\newcommand{\posGd}{\ensuremath{-0.6951890756789}} 

\newcommand{\FstarzeroCasA}{\ensuremath{65.8} }
\newcommand{\FstarzeroVela}{\ensuremath{52.6} }
\newcommand{\FstarzeroG}{\ensuremath{45.4} }

\newcommand{\paramfdothiCasA}{\ensuremath{0}} 
\newcommand{\paramfdotloCasA}{\ensuremath{-\sci{9.6}{-9}}}
\newcommand{\paramfdotloVela}{\ensuremath{-\sci{4.5}{-9}}} 
\newcommand{\paramfdothiVela}{\ensuremath{0}}
\newcommand{\paramfdothiG}{\ensuremath{0}} 
\newcommand{\paramfdotloG}{\ensuremath{-\sci{2.0}{-9}}} 
\newcommand{\paramfddothiCasA}{\ensuremath{\sci{4.6}{-18}}} 
\newcommand{\paramfddotloCasA}{\ensuremath{0}}
\newcommand{\paramfddotloVela}{\ensuremath{0}} 
\newcommand{\paramfddothiVela}{\ensuremath{\sci{1.0}{-18}}}
\newcommand{\paramfddothiG}{\ensuremath{\sci{2.0}{-19}}}
\newcommand{\paramfddotloG}{\ensuremath{0}} 

\newcommand{\ageCasA}{\ensuremath{330} }
\newcommand{\ageVela}{\ensuremath{700} }
\newcommand{\ageG}{\ensuremath{1600}}

\newcommand{\TrefGPS}{\ensuremath{1131943508}}

\newcommand{\paramWUcputimeHours}{8} 
\newcommand{\paramtotalWUsmillions}{3} 
\newcommand{\paramtotaltemplates}{\ensuremath{\sci{2}{17}}} 

\newcommand{\ThrVelalow}{6.2}
\newcommand{\ThrVelamid}{6.9}
\newcommand{\ThrVelahigh}{8.3}

\newcommand{\ThrCasAlow}{5.3}
\newcommand{\ThrCasAmid}{7.6}
\newcommand{\ThrCasAhigh}{8.3}

\newcommand{\ThrGlow}{5.9}
\newcommand{\ThrGmid}{6.2}
\newcommand{\ThrGhigh}{7.7}

\newcommand{\ClustNrangeVelalow}{[150, 225]}
\newcommand{\ClustNrangeVelamid}{[150, 225]}
\newcommand{\ClustNrangeVelahigh}{[150, 225]}

\newcommand{\ClustCVelalow}{1000}
\newcommand{\ClustCVelamid}{1000}
\newcommand{\ClustCVelahigh}{100}

\newcommand{\ClustPVelalow}{0.25}
\newcommand{\ClustPVelamid}{0.25}
\newcommand{\ClustPVelahigh}{0.25}

\newcommand{\ClustDVelalow}{0.95}
\newcommand{\ClustDVelamid}{0.9}
\newcommand{\ClustDVelahigh}{0.9}

\newcommand{\ClustGVelalow}{0.0001}
\newcommand{\ClustGVelamid}{0.01}
\newcommand{\ClustGVelahigh}{0.01}

\newcommand{\ClustNrangeCasAlow}{[75, 150]}
\newcommand{\ClustNrangeCasAmid}{[75, 150]}
\newcommand{\ClustNrangeCasAhigh}{[75, 150]}

\newcommand{\ClustCCasAlow}{10}
\newcommand{\ClustCCasAmid}{10}
\newcommand{\ClustCCasAhigh}{10}

\newcommand{\ClustPCasAlow}{0.25}
\newcommand{\ClustPCasAmid}{0.25}
\newcommand{\ClustPCasAhigh}{0.25}

\newcommand{\ClustDCasAlow}{0.8}
\newcommand{\ClustDCasAmid}{0.8}
\newcommand{\ClustDCasAhigh}{0.9}

\newcommand{\ClustGCasAlow}{1.0}
\newcommand{\ClustGCasAmid}{1.0}
\newcommand{\ClustGCasAhigh}{1.0}

\newcommand{\ClustNrangeGlow}{[125, 250]}
\newcommand{\ClustNrangeGmid}{[125, 250]}
\newcommand{\ClustNrangeGhigh}{[125, 250]}

\newcommand{\ClustCGlow}{300}
\newcommand{\ClustCGmid}{400}
\newcommand{\ClustCGhigh}{200}

\newcommand{\ClustPGlow}{0.3}
\newcommand{\ClustPGmid}{0.3}
\newcommand{\ClustPGhigh}{0.3}

\newcommand{\ClustDGlow}{0.98}
\newcommand{\ClustDGmid}{0.95}
\newcommand{\ClustDGhigh}{0.95}

\newcommand{\ClustGGlow}{0.001}
\newcommand{\ClustGGmid}{0.01}
\newcommand{\ClustGGhigh}{0.1}

\newcommand{\NcandVelatot}{40}

\newcommand{\NcandCasAtot}{21}

\newcommand{\NcandGtot}{23}

\newcommand{\NrejectDMoffVela}{8}
\newcommand{\NrejectDMoffCasA}{14}
\newcommand{\NrejectDMoffG}{2}

\newcommand{\RoneC}{3.5}
\newcommand{\RoneV}{2.8}
\newcommand{\RoneG}{2.1}

\newcommand{\NrejectFUoneVela}{38}
\newcommand{\NrejectFUoneCasA}{18}
\newcommand{\NrejectFUoneG}{21}

\newcommand{\RtwoC}{7}
\newcommand{\RtwoV}{5}
\newcommand{\RtwoG}{4}

\newcommand{\lowestULc}{$1.3\times 10^{-25}$}

\newcommand{\lowestULv}{$1.0\times 10^{-25}$}

\newcommand{\lowestULg}{$9.5\times 10^{-26}$}

\newcommand\Tstrut{\rule{0pt}{2.9ex}}       
\newcommand\Bstrut{\rule[-1.3ex]{0pt}{0pt}} 
\newcommand\TBstrut{\Tstrut\Bstrut}  

\newtoggle{fullauthorlist}
\toggletrue{fullauthorlist}

\newtoggle{endauthorlist}
\toggletrue{endauthorlist}

\begin{document}

\title{ 
Results from an Einstein@Home search for continuous gravitational waves from Cassiopeia A, Vela Jr. and G347.3
}


\author{J. Ming}
\email{jing.ming@aei.mpg.de}
\affiliation{Max Planck Institute for Gravitational Physics (Albert Einstein Institute), Callinstrasse 38, 30167 Hannover, Germany}
\affiliation{Leibniz Universit\"at Hannover, D-30167 Hannover, Germany}

\author{M.A. Papa}
\email{maria.alessandra.papa@aei.mpg.de}
\affiliation{Max Planck Institute for Gravitational Physics (Albert Einstein Institute), Callinstrasse 38, 30167 Hannover, Germany}
\affiliation{University of Wisconsin Milwaukee, 3135 N Maryland Ave, Milwaukee, WI 53211, USA}
\affiliation{Leibniz Universit\"at Hannover, D-30167 Hannover, Germany}

\author{A. Singh}
\affiliation{Max Planck Institute for Gravitational Physics (Albert Einstein Institute), Callinstrasse 38, 30167 Hannover, Germany}
\affiliation{Leibniz Universit\"at Hannover, D-30167 Hannover, Germany}
\affiliation{The Geophysical Institute, Bjerknes Centre for Climate Research, University of Bergen, 5007 Bergen, Norway}

\author{H.-B. Eggenstein}
\affiliation{Max Planck Institute for Gravitational Physics (Albert Einstein Institute), Callinstrasse 38, 30167 Hannover, Germany}
\affiliation{Leibniz Universit\"at Hannover, D-30167 Hannover, Germany}

\author{S. J. Zhu}
\affiliation{Max Planck Institute for Gravitational Physics (Albert Einstein Institute), Callinstrasse 38, 30167 Hannover, Germany}
\affiliation{Leibniz Universit\"at Hannover, D-30167 Hannover, Germany}

\author{V. Dergachev}
\affiliation{Max Planck Institute for Gravitational Physics (Albert Einstein Institute), Callinstrasse 38, 30167 Hannover, Germany}
\affiliation{Leibniz Universit\"at Hannover, D-30167 Hannover, Germany}

\author{Y. Hu}
\affiliation{Max Planck Institute for Gravitational Physics (Albert Einstein Institute), Callinstrasse 38, 30167 Hannover, Germany}
\affiliation{Leibniz Universit\"at Hannover, D-30167 Hannover, Germany}
\affiliation{TianQin Research Center for Gravitational Physics, Sun Yat-Sen University, SYSU Zhuhai Campus, Tangjiawan, Zhuhai 519082, Guangdong, P. R. China}

\author{R. Prix}
\affiliation{Max Planck Institute for Gravitational Physics (Albert Einstein Institute), Callinstrasse 38, 30167 Hannover, Germany}
\affiliation{Leibniz Universit\"at Hannover, D-30167 Hannover, Germany}

\author{B. Machenschalk}
\affiliation{Max Planck Institute for Gravitational Physics (Albert Einstein Institute), Callinstrasse 38, 30167 Hannover, Germany}
\affiliation{Leibniz Universit\"at Hannover, D-30167 Hannover, Germany}

\author{C. Beer}
\affiliation{Max Planck Institute for Gravitational Physics (Albert Einstein Institute), Callinstrasse 38, 30167 Hannover, Germany}
\affiliation{Leibniz Universit\"at Hannover, D-30167 Hannover, Germany}
\affiliation{Rechenkraft.net e.V., 35039 Marburg, Germany}

\author{O. Behnke}
\affiliation{Max Planck Institute for Gravitational Physics (Albert Einstein Institute), Callinstrasse 38, 30167 Hannover, Germany}
\affiliation{Leibniz Universit\"at Hannover, D-30167 Hannover, Germany}

\author{B. Allen}
\affiliation{Max Planck Institute for Gravitational Physics (Albert Einstein Institute), Callinstrasse 38, 30167 Hannover, Germany}
\affiliation{University of Wisconsin Milwaukee, 3135 N Maryland Ave, Milwaukee, WI 53211, USA}
\affiliation{Leibniz Universit\"at Hannover, D-30167 Hannover, Germany}

\begin{abstract}
We report results of the most sensitive search to date for periodic gravitational waves from Cassiopeia A, Vela Jr. and G347.3 with frequency between 20 and 1500 Hz. The search was made possible by the computing power provided by the volunteers of the \EatHs project and improves on previous results by a factor of 2 across the entire frequency range for all targets. We find no significant signal candidate and set the most stringent upper limits to date on the amplitude of gravitational wave signals from the target population, corresponding to sensitivity depths between 54 $[1/ {\sqrt{\textrm{Hz}}}]$ and 83 $[1/ {\sqrt{\textrm{Hz}}}]$, depending on the target and the frequency range. At the frequency of best strain sensitivity, near $172$ Hz, we set 90\%\ confidence upper limits on the gravitational wave intrinsic amplitude of $h_0^{90\%}\approx 10^{-25}$, probing ellipticity values for Vela Jr. as low as $3\times 10^{-8}$, assuming a distance of 200 pc. 

\end{abstract}

\maketitle

\section{Introduction}
\label{sec:introduction}

A continuous stream of weak gravitational waves is expected when a compact spinning object such as a neutron star presents deviations from an axisymmetric configuration. The simplest form of non-axisymmetric configuration is when the star is deformed; The deformation is usually expressed in terms of the equatorial ellipticity of the object $\varepsilon ={|{I_{xx}-I_{yy}|}\over {I_{zz}}}$, with $I$ being its moment of inertia tensor. 
Compact objects of normal baryonic matter could sustain ellipticities of up to $10^{-5}$ and mechanisms have been proposed for processes to produce such deformations \cite{McDanielJohnsonOwen}.

Newly born compact stars are likely to have large deformations and be spinning rapidly. For this reason young neutron star candidates are considered interesting targets for continuous wave emission. Several directed searches have been performed in the past, targeting supernova remnants in search of continuous gravitational wave emission from the putative young compact object that the remnant may harbour \cite{Abbott:2018qee,S6EHCasA,Aasi:2014ksa}. 

Since no pulsations are observed from these objects the range of possible signal frequency and spindown values is very broad, and the deepest searches are very computationally intensive. For instance, the search \cite{S6EHCasA} performed on Initial LIGO data, took months on the volunteer computing project \EatH.

An optimisation scheme has been proposed to rationally decide how to spend the available computational budget, optimally distributing the resources among the different targets and the parameter space, in such a way to maximise the probability of making a gravitational wave detection \cite{Ming:2015jla}. The scheme forces one to make explicit all the assumptions on source parameters and consistently consider them. Applying the procedure to a set of interesting point sources, likely compact objects at the center of young supernova remnants, with a computational budget of a few months on \EatH, yields a specific set-up for a search on data from the first Advanced LIGO run (O1) \cite{thankslosc,O1data} targeting Vela Jr. (G266.2-1.2), Cassiopeia A (G111.7-2.1) and G347.3 (G347.3-0.5) \cite{Ming:2017anf}.

We carry out the search on the \EatH~ infrastructure, the post-processing of the results on the in-house super-computing cluster  Atlas \cite{ATLAS} and we describe the results in this paper. 

The paper is organized as follows: after a brief description of the data in Section \ref{sec:Data}, we summarise the primary search run on \EatH~ in Section \ref{sec:search} and the hierarchical follow-up searches in Section \ref{sec:followup}. The results follow in Section \ref{sec:results}. There we explain how the $h_0^{90\%}$ upper limits on the intrinsic continuous gravitational wave strain amplitude are computed and how the ellipticity and r-mode amplitude upper limits are derived from these. We conclude with a discussion of the results, comparing and contrasting with existing literature in Section \ref{sec:conclusions}.

\section{LIGO interferometers and the data used}
\label{sec:Data} 

The LIGO gravitational wave network consists of two observatories in the USA, one in Hanford (WA) and the other in Livingston (LA) \cite{LIGO_detector}. The O1 run of this network, famous for the first direct gravitational wave detection, took place between September 2015 and January 2016 \cite{LIGO_O1,TheLIGOScientific:2014jea,o1_data,aligo}. We use data from this run.

Due to environmental or instrumental disturbances or because of scheduled maintenance periods, each detector has a duty factor of about 50\% and the data set that can be used for scientific analyses is not continuous.  Fourier transforms of data segments 1800s long (SFTs) are created \cite{SFTs}. To remove the effects of instrumental and environmental spectral disturbances from the analysis, the data in frequency bins known to contain such disturbances is substituted with Gaussian noise with the same average power as that in the neighbouring and undisturbed bands. This is the same procedure as used in \cite{Abbott:2017pqa} and previous \EatH~ runs. This  SFT data constitute the input to the search.

\section{The Search}
\label{sec:search}

The search described in this paper targets nearly monochromatic gravitational wave signals of the form described for example in Section II of  \cite{Jaranowski:1998qm} from the three supernova remnants Cassiopeia A (Cas A, hereafter), Vela Jr. and G347.3.  

We perform a stack-slide type of search using the GCT (Global correlation transform) method \cite{PletschAllen,Pletsch:2008,Pletsch:2010}. In a stack-slide search the data is partitioned in segments, and each segment is searched with a maximum likelihood multi-detector coherent method \cite{Cutler:2005hc}, the so-called $\F$-statisitic. The results from these coherent searches are combined by summing the detection statistic values from the different segments, one per segment ($\F_i$), and this determines the value of the core detection statistic: 
\begin{equation}
\label{eq:avF}
\avF:={1\over\Nseg} \sum_{i=1}^{\Nseg} \F_i.
\end{equation}
The ``stacking'' part of the procedure is the summing and the ``sliding'' (in parameter space) refers to the fact that the $\F_i$ that are summed do not all come from the same template. 

In order to ease the impact of coherent disturbances present in the data, from the multi-detector and single-detector coherent detection statistics, 
we also compute the transient-line-robust detection statistics $\BSNtsc$ \cite{Keitel:2013,Keitel:2016}. We use tuning parameter $\scFtho =$ \FstarzeroCasA, \FstarzeroVela and \FstarzeroG for Cas A, Vela Jr. and G347.3 respectively and equal-odds priors between the various noise hypotheses (``L'' for line, ``G" for Gaussian, ``tL" for transient-line). 
The $\scFtho$ values given above have been computed using Eq.~67 of~\cite{Keitel:2013} to yield a Gaussian false-alarm probability of $10^{-9}$ for respective searches.


Important variables for a stack-slide search are: the coherent time baseline of the segments $\Tcoh$, the number of segments used $\Nseg$, the total time $\Tobs$ spanned by the data, the grids in parameter space and the detection statistic used to rank the parameter space cells. These parameters are given in Table \ref{tab:GridSpacings}.  
For a stack-slide search in Gaussian noise, $\Nseg\times 2\avF$ follows a chi-squared distribution with $4\Nseg$ degrees of freedom, $\chi^2_{4\Nseg}$.

The search parameters follow the prescription given in \cite{Ming:2017anf}, with search ranges in spin-down defined as follows: 
\begin{equation}
\label{eq:Priors}
	\begin{cases}
	-f/ \tau\, \le   \dot{f} ~\le 0\,~\mathrm{Hz/s}\\
	0\,\mathrm{Hz/s}^2 \leq  \ddot{f} \leq ~5\dot{|f|}_{\textrm{max}}^2/f = 5 {f/\tau^2}, 
	\end{cases}
\end{equation}
and $\tau=$ $\ageCasA$ yrs, $\ageVela$ yrs, $\ageG$ yrs for Cas A, Vela Jr. and G347.3 respectively. Table \ref{tab:SearchParams} shows the numeric values of the ranges for $f=100 Hz$. 

The grids in frequency and spindown are each described by a single parameter, the grid spacing, which is constant over the search range. The same grid spacings are used for frequency both in the coherent searches over the segments and in the incoherent summing. The first and second order spin-down spacings for the incoherent summing, $\delta{\dot{f}}, \delta{\ddot{f}}$, are finer than those used for the coherent searches, $\delta{\dot{f_c}}, \delta{\ddot{f_c}}$, by factors $\gamma_1$ and $\gamma_2$ respectively. The measured average mismatch for the chosen grids for the Cas A, Vela Jr. and G347.3 searches is 41\%, 16\% and 12\%, respectively \cite{Ming:2017anf}.

\begin{figure}[h!tbp]
  \includegraphics[width=\columnwidth]{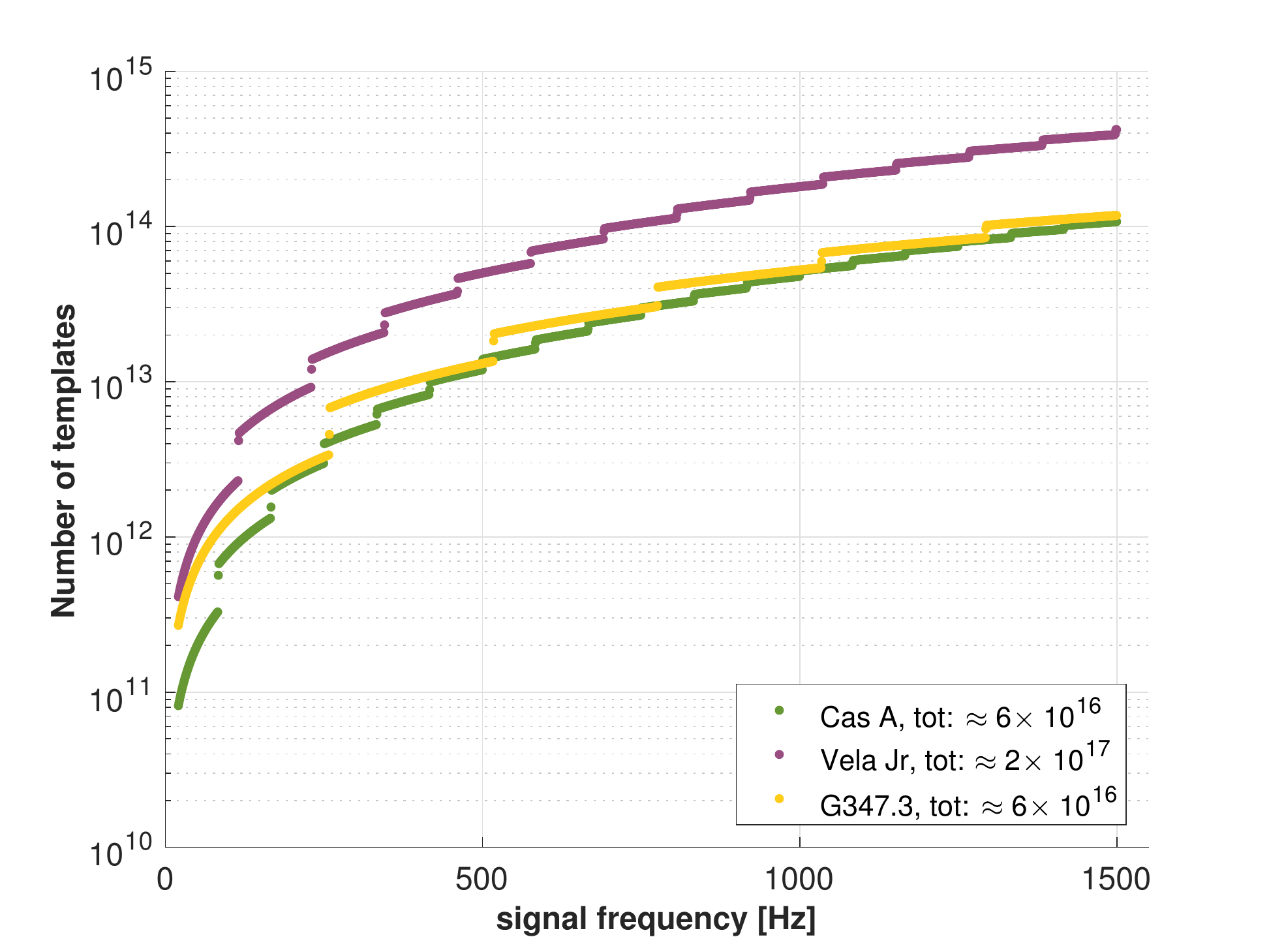}
\caption{Number of templates searched in 1-Hz bands as a function of signal frequency for the different targets. In the legend we also show the total number of templates searched for each target.}
\label{fig:HowManyTemplates}
\end{figure}

The number of templates searched in a given frequency interval varies as a function of frequency and from target to target.  The reason is that the grid spacings are different for the different targets and the searched spindown ranges grow with frequency and decrease with increasing age of the target. Fig. \ref{fig:HowManyTemplates} shows the number of templates searched in 1-Hz bands as a function of frequency for the three targets. 

\begin{table*}[ht]
\begin{tabular}{|c|c|c|c|}
\hline
\hline
 & Cas A & Vela Jr. & G347.3 \\
\hline
\hline
 \TBstrut$f$ range & \multicolumn{3}{|c|} {[20-1500] Hz} \\
 \hline
  \TBstrut$\Tref$  & \multicolumn{3}{|c|} {\TrefGPS GPS s } \\
  \hline
\TBstrut$\fdot $ range &  [\paramfdotloCasA - \paramfdothiCasA~ ] Hz/s & [\paramfdotloVela - \paramfdothiVela~ ]  Hz/s & [\paramfdotloG - \paramfdothiG ~] Hz/s \\
 \hline
 \TBstrut$\fddot $ range &  [ \paramfddotloCasA~  -  \paramfddothiCasA~ ] Hz/s$^2$  & [ \paramfddotloVela~ - \paramfddothiVela ~] Hz/s$^2$   & [ \paramfddotloG~ - \paramfddothiG~ ] Hz/s$^2$  \\
  \hline
\TBstrut$ \alpha$ & \posCasAa & \posVelaJra & \posGa \\  
\TBstrut$ \delta$  &  \posCasAd & \posVelaJrd & \posGd \\
 \hline
\hline
\end{tabular}
\caption{Search ranges. The spindown ranges quoted are the ones used at 100 Hz. The ranges at different frequencies are readily derived from Eq.~\ref{eq:Priors}. }
\label{tab:SearchParams}

\begin{tabular}{|c|c|c|c|}
\hline
\hline
 & Cas A & Vela Jr. & G347.3 \\
\hline
\hline
 \TBstrut$\Tcoh$ & 245 hr & 369 hr& 489 hr\\
 \hline
\TBstrut$\Nseg$ & 12 & 8 & 6 \\
  \hline
\TBstrut$\delta f$ & $6.85 \times 10^{-7}$ Hz & $3.21 \times 10^{-7}$ Hz & $2.43 \times 10^{-7}$ Hz \\
 \hline
 \TBstrut$\delta {\dot{f_c}}$ &  $3.88\times 10^{-12}$ Hz/s  & $1.33\times 10^{-12}$ Hz/s  & $6.16 \times 10^{-13}$ Hz/s \\
  \hline
\TBstrut$\gamma_1$ & 5 &  9 & 9 \\
  \hline
\TBstrut$\delta {\ddot{f_c}}$ & $ 4.03\times 10^{-18} {\textrm{Hz/s}}^2 $ & $1.18\times 10^{-18} {\textrm{Hz/s}}^2 $ & $5.07 \times 10^{-19} {\textrm{Hz/s}}^2$ \\
  \hline
\TBstrut$\gamma_2$ & 21 & 21 & 11 \\
 \hline
\hline  
\end{tabular}
\caption{Spacings on the signal parameters used for the templates in the \EatH~ search. }
\label{tab:GridSpacings}
\end{table*}

This search was performed on the \EatH~volunteer computing project. \EatH~  is built on the BOINC (Berkeley Open Infrastructure for Network Computing) architecture~\cite{Boinc1,Boinc2,Boinc3} which uses the idle time on volunteer computers to tackle scientific problems such as this, that require large amounts of computer power. Overall about \paramtotaltemplates~ templates were searched. The computational work was split into work-units (WUs) sized to keep the average \EatHs volunteer computer busy for about {\paramWUcputimeHours} CPU-hours. {\paramtotalWUsmillions}~million WUs were computed to carry out this search, not including redundancy for cross-validation.  Only the highest 10000 detection statistic values in each WU were communicated back to the central \EatH~ server.

\subsection{Identification of undisturbed bands}
\label{sec:visualInspection}

Even after the removal of disturbed data caused by spectral artefacts of known origin, the statistical properties of the results are not uniform across the search band. In what follows we concentrate on the subset of the signal-frequency bands having reasonably uniform statistical properties, or containing features that are not immediately identifiable as detector artefacts. This comprises the large majority of the search parameter space.

Our classification of ``clean" vs. ``disturbed" bands has no pretence of being strictly rigorous, because strict rigour here is neither useful nor practical. The classification serves the practical purpose of discarding from the analysis regions in parameter space with evident disturbances and must not dismiss detectable real signals. 

An automatic procedure, described in Section IIF of \cite{S6EHCasA}, identifies as undisturbed the 50-mHz bands whose maximum density of outliers in the $f-\dot{f}$ plane and average $2\avF$ are well within the bulk distribution of the values for these quantities in the neighbouring frequency bands. This procedure identifies $\lesssim 3\%$ of the bands as potentially disturbed, with a much higher concentration (greater by a factor of $\approx$ 5) of disturbed bands below 100 Hz.

We use the line-robust and transient-line-robust $\BSNtsc$, which we recalculate exactly at the parameter space point for which the \EatH~ result provides an approximate value. We indicate this recalculated value by $\BSGLtLr$. The detection threshold is constant in the frequency ranges  [20-250] Hz , [250-520] Hz and [520-1500] Hz. We refer to these as the low, mid and high frequency range, respectively.

\begin{table}[t]
\begin{tabular}{|c|c|c|c|}
\hline
\hline
 \TBstrut& Cas A & Vela Jr. & G347.3 \\
\hline
\hline
 \TBstrut$~\BSGLtLr^{\texttt{low-freq}}~$ &  $\ThrCasAlow$ & $\ThrVelalow$  & $\ThrGlow$ \\
 \hline
\TBstrut$~\BSGLtLr^{\texttt{mid-freq}}~$ &  $\ThrCasAmid$ & $\ThrVelamid$  & $\ThrGmid$ \\
 \hline
\TBstrut$~\BSGLtLr^{\texttt{high-freq}}~$ &  $\ThrCasAhigh$ & $\ThrVelahigh$  & $\ThrGhigh$ \\
 \hline
\hline  

\end{tabular}
\caption{Stage 0 detection statistic thresholds.}
\label{tab:HighThresholds}
\end{table}

We pick high thresholds (see table \ref{tab:HighThresholds}) which results in computationally very light follow-up stages. A deep search with lower thresholds would be much more demanding computationally and at the end requires significant follow-ups using a different data set (see for example \cite{Aasi:2015rar, Dergachev:2019pgs}). Since the searches reported here were performed before the release of the LIGO O2 data set, not having access to a follow-up data set, we concentrated on highly significant candidates. Lower significance candidates will be pursued in a future paper. We call ``candidates'' all detection statistic values, and associated waveform parameters, above the thresholds.

\section{Hierarchical follow Up}
\label{sec:followup}

We investigate the  candidates above threshold to determine if they are produced by a signal or by a detector disturbance. This is done with a hierarchical approach similar to \cite{S6EHFU, Abbott:2017pqa} or more recently \cite{Dergachev:2019pgs,O2AllSkyLVC}. 

At each stage of the hierarchical follow-up a semi-coherent search is performed, the top ranking candidates are marked and then searched in the next stage. If the data harbours a real signal, the significance of the recovered candidate will increase with respect to the significance that it had in the previous stage. On the other hand, if the candidate is not produced by a continuous-wave signal, the significance is not expected to increase consistently over the successive stages. The status of each candidate through the follow-up stages is shown in figure \ref{fig:CandsStage0}.

The hierarchical approach used in this search consists of three stages. 

\subsection{Stage 0}
\label{sec:stage0}

\begin{figure}[h!]
    \includegraphics[width=1 \columnwidth]{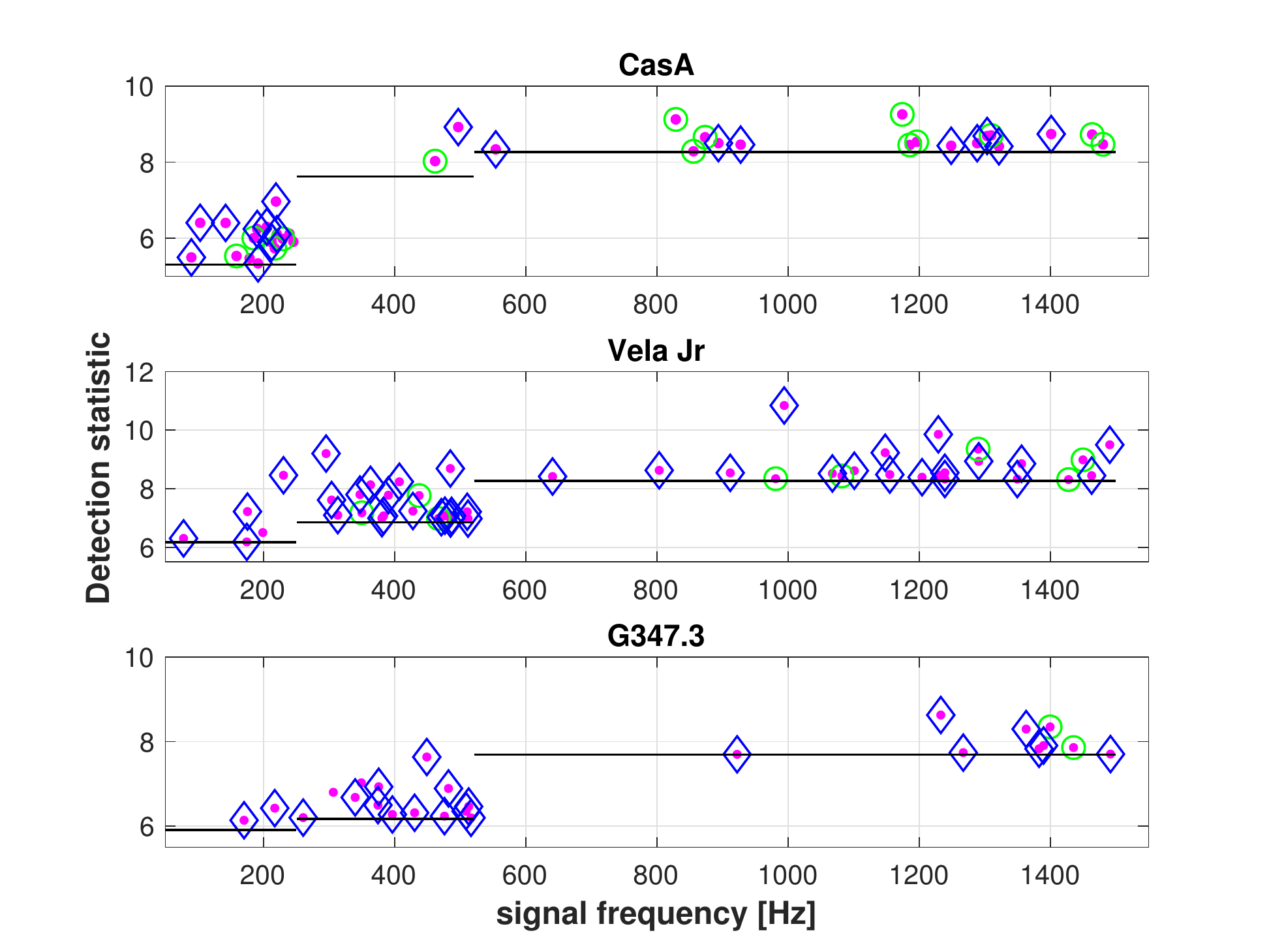}
    \caption{Detection statistic $\BSGLtLr$ as a function of signal frequency of candidates. The smaller filled circles (magenta) are the candidates above the Stage 0 threshold. There are \NcandCasAtot, \NcandVelatot~ and \NcandGtot~candidates for Cas A, Vela Jr. and G347.3, respectively. The larger (green) circles indicate the \NrejectDMoffCasA, \NrejectDMoffVela~and \NrejectDMoffG~ candidates that do not pass the DM-off veto from Cas A, Vela Jr. and G347.3, respectively. The diamonds (purple) show the  \NrejectFUoneCasA, \NrejectFUoneVela~ and \NrejectFUoneG~  candidates that are rejected after the first stage follow-up.}
\label{fig:CandsStage0}
\end{figure}

A clustering procedure \cite{Singh:2017kss} identifies as due to the same root-cause, close-by candidates in parameter space. The clustering parameters are given in the Appendix \ref{sec:clusteringParams}. 
After clustering we have \NcandCasAtot~ candidates from Cas A, \NcandVelatot~ candidates from Vela Jr.  and \NcandGtot~ candidates from G347.3 above threshold. Figure \ref{fig:CandsStage0} shows the detection statistic values of these candidates and their template frequencies. 

A semi-coherent DM-off veto \cite{Zhu:2017ujz,DMoff-sc} is applied to these candidates. This veto is based on a comparison between the detection statistic value obtained in the original astrophysical search and the detection statistic value that is obtained in a search for non-astrophysical signals, i.e. signals that do not present any Doppler modulation. The veto is tuned to be safe and computationally feasible. The noise rejection is assessed a-posteriori on the data. In this case the veto is not as effective as in the past \cite{Abbott:2017pqa}, rejecting only \NrejectDMoffVela, \NrejectDMoffCasA~ and \NrejectDMoffG~ candidates for Vela Jr., Cas A and G347.3, respectively. 

\subsection{Stage 1}
\label{sec:stage1}

We follow-up each candidate that survives the previous stage with a semi-coherent search with a coherent time-baseline, of approximately 60 days.  We fix the total run time to be $\sim 8$ hrs on a few hundred nodes of the ATLAS computing cluster \cite{ATLAS}. The set-up that minimizes the average mismatch, at $\sim 9\%$, for this computing budget has the following grid spacings: 
\begin{equation}
\begin{cases}
\delta f^{\textrm{(1)}} & \simeq 8\times 10^{-8}\, {\textrm{Hz}}\\
\delta \dot{f}^{\textrm{(1)}} &\simeq  1\times 10^{-14}\, {\textrm{Hz/s}}\\
\delta \ddot{f}^{\textrm{(1)}} &\simeq  2\times 10^{-21}\, {\textrm{Hz/s}}^2,
\end{cases}
\label{eq:FU1Grids}
\end{equation}
where the superscript ``(1)" indicates that this is the first follow-up stage. 

From each search we take the most significant candidate as measured by the $\BSGLtLr$.
We set a threshold on the ratio $R$ defined as 
\begin{equation}
R^{\textrm{(1)}}:={{2\avF^{\textrm{(1)}} - 4 } \over {{2\avF^{\textrm{(0)}} - 4}}},
\label{eq:Rdef}
\end{equation}
where the superscript ``0'' indicates that the detection statistic value comes from the original \EatH~ search. With Monte Carlos of simulated follow-up results containing fake signals we experiment in constructing different discriminators, also using the transient- and line-robust statistic $\BSNtsc$ but find that Eq.~\ref{eq:Rdef} defines the most efficient detection statistic to identify the candidates for the next stage. With ``most efficient'' here we mean that at fixed false dismissal it yields the lowest false alarm. We use the following thresholds on $R^{(1)}$ for Cas A, Vela Jr., and G347.3 respectively: \RoneC, ~\RoneV ~and \RoneG. \NrejectFUoneCasA, ~ \NrejectFUoneVela ~and \NrejectFUoneG ~candidates are rejected (see Figure \ref{fig:CandsStage0}).

\subsection{Stage 2}
\label{sec:stage2}

In this stage we follow up the remaining 3, 2 and 2 candidates from Cas A, Vela Jr. and G347.3. We use a fully coherent search over the entire data set. The search set up has an average mismatch of 0.34\% and the following grid spacings:
\begin{equation}
\begin{cases}
\delta f^{\textrm{(1)}} & \simeq 2\times 10^{-8}\, {\textrm{Hz}}\\
\delta \dot{f}^{\textrm{(1)}} &\simeq  5\times 10^{-15}\, {\textrm{Hz/s}}\\
\delta \ddot{f}^{\textrm{(1)}} &\simeq  2\times 10^{-21}\, {\textrm{Hz/s}}^2.
\end{cases}
\label{eq:FU2Grids}
\end{equation}
We use the same procedure as for Stage 1, with threshold values on $R^{\textrm{(2)}}$ of \RtwoC, \RtwoV~ and \RtwoG~ respectively for Cas A, Vela Jr. and G347.3. $R^{\textrm{(2)}}$ is the quantity defined in Eq. \ref{eq:Rdef}, with the superscript ``(2)'' indicating that this is the second follow-up stage. No candidate passes this threshold.

We conclude that it is unlikely that any of our candidates arises from a long-lived astronomical source of continuous gravitational waves. We proceed to set upper limits on the amplitude of such signals from the three targets.

\section{Results}
\label{sec:results}

\subsection{Upper limits on the gravitational wave amplitude}

The search does not reveal any continuous gravitational wave signal hence we set frequentist 90\% confidence upper limits on the maximum gravitational wave amplitude consistent with this null result as function of the signal frequency,  $h_0^{90\%}(f)$. Specifically, $h_0^{90\%}(f)$ is the GW amplitude such that 90\% of a population of signals with parameter values in our search range would have been detected by our search. 
\begin{figure*}
   \includegraphics[width=0.85\textwidth]{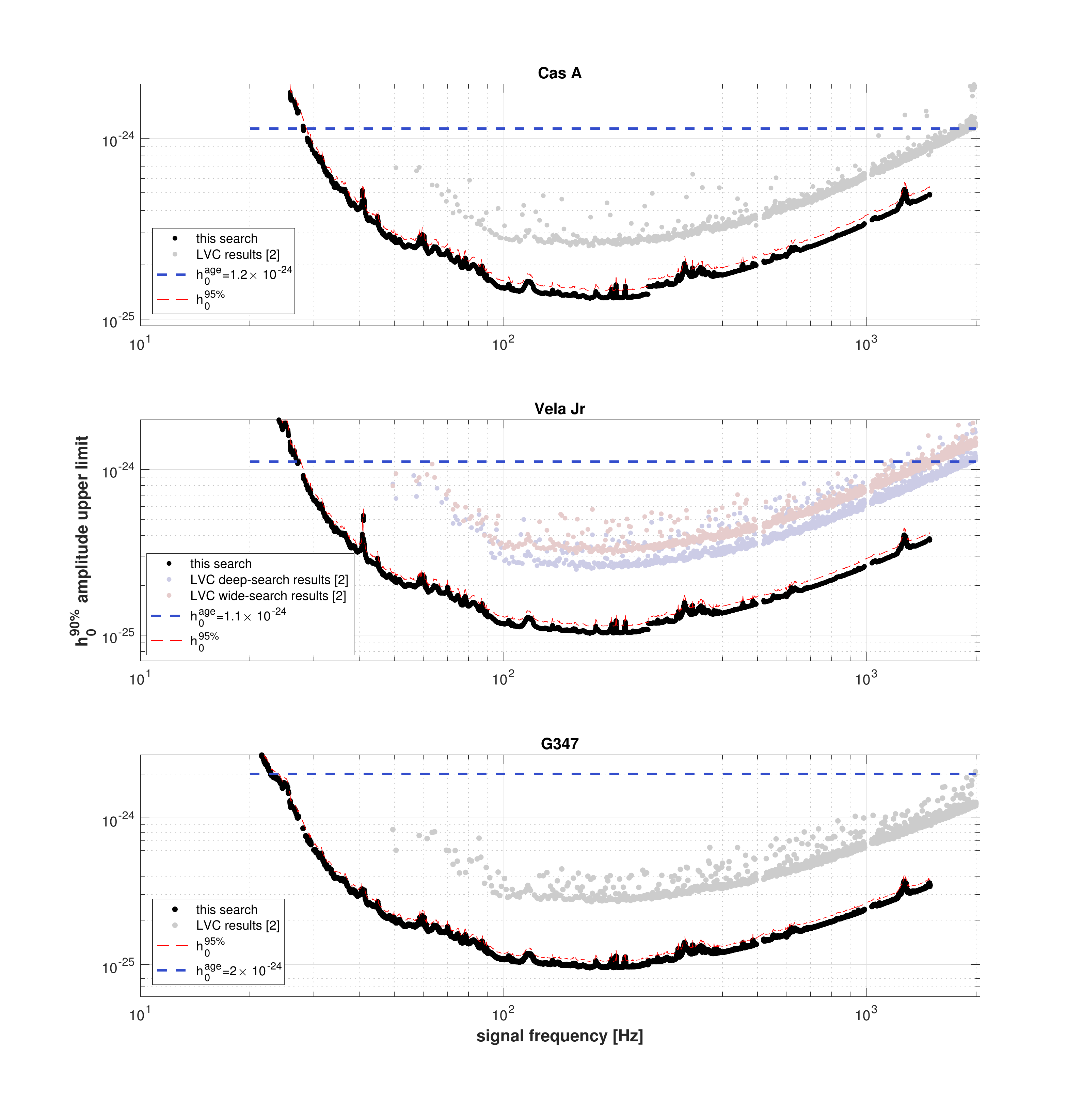}
\caption{90\% confidence upper limits on the gravitational wave amplitude of continuous gravitational wave signals for the three targets with spindown values within the searched ranges (\ref{eq:Priors}) as a function of signal frequency. The lowest set of points (black circles) are the results of this search. For comparison we also show in lighter shades recent upper limits results for these targets from the LIGO/Virgo \cite{Abbott:2018qee}. For Vela Jr. \cite{Abbott:2018qee} perform two different searches: one ``deep'', more sensitive search, assuming that the star is old and far, and the other, a search wider primarily in spindown, assuming the star to be young and close. Since the upper limits from \cite{Abbott:2018qee} are at 95\% confidence, for completeness the dashed (red) curves show our results for 95\% confidence. The (blue) dashed line at the top shows the age-based upper limit. For Vela Jr. we show the most constraining age limit, i.e. the one assuming the object is farther away (900 pc) and older (5100 yrs). The limit under the assumption that Vela Jr. is young (700 yrs) and close-by (200 pc) is $1.4\times 10^{-23}$.
}  
\label{fig:ULs}
\end{figure*}

Since an actual full scale fake-signal search-and-recovery Monte Carlo for the entire 1480 Hz search range is prohibitive, in the same spirit as \cite{Abbott:2017pqa, S6EHCasA, S6EHFU, AvneetThesis}, we perform such a study in a limited set of trial bands. We choose the following 2-Hz bands in the search range to measure the upper limits:  101.9-103.9 Hz , 202-204 Hz,  339-341 Hz, 443-445 Hz, 684-686 Hz, 1054-1056 Hz , 1404-1406 Hz . All these bands were marked as undisturbed by the automated classification scheme that we used.

\begin{figure}[h!tbp]
   \includegraphics[width=\columnwidth]{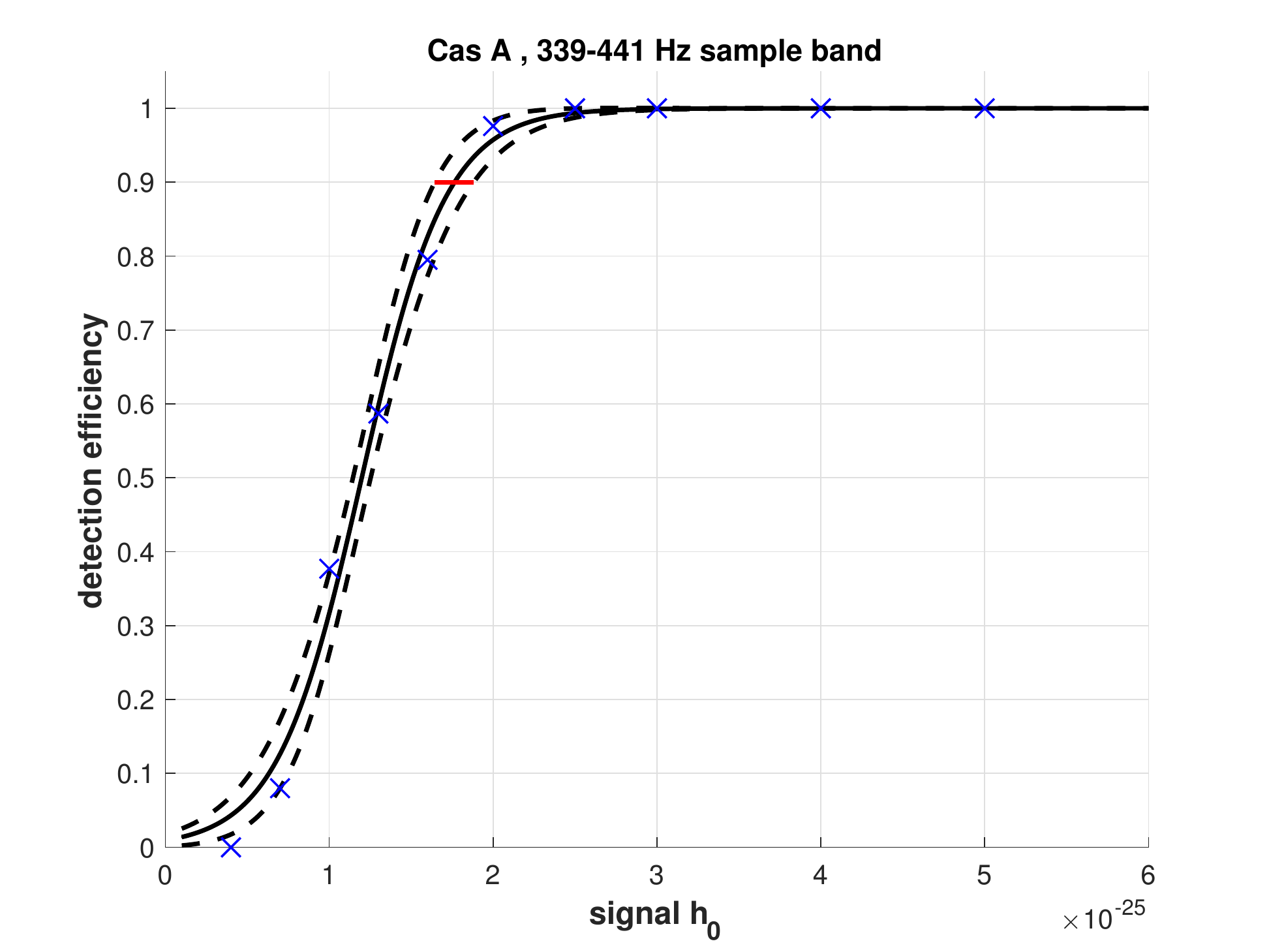}
\caption{Blue crosses: measured detection rate $C(h_0)$ from the CasA signal search-and-recovery Monte Carlos with signal frequencies between 339 and 441 Hz. The solid line is the best fit and the dashed lines represent 95\% confidence intervals on the fit. The red line marks the 90\% detection confidence level.}
\label{fig:exampleSigmoid}
\end{figure}
We simulate 1000 signals in the real detector data at fixed value of the intrinsic amplitude $h_0$, drawing all other signal parameters -- the frequency, inclination angle $\cos\iota$, polarization $\psi$ and initial phase values --  from a uniform random distribution in their respective ranges. The spindown values are log-uniform within the search range at each frequency. We consider 10 values of $h_0$ , spanning the range $[4\times 10^{-26}-5\times 10^{-25}]$. A search is performed with the same grids and set-up as the original \EatH~ search, in the neighbourhood of the fake signal parameters. Since the follow-up stages have a very small false dismissal rate, a signal is considered recovered if a detection statistic value is found in the search results above the Stage-0 detection threshold. Counting the fraction of recovered signals out of the total 1000, yields a detection rate, or detection confidence, for each value of $h_0$. The $h_0$ versus confidence data is fit with a sigmoid of the form 
\begin{equation}
C(h_0)={1\over{1+\exp({{{\textrm{a}}-h_0}\over{\textrm{b}}})}}
\label{eq:sigmoidFit}
\end{equation}
and the $h_0^{90\%}$ value is read-off of this curve. An associated 95\% credible interval on the fit is also derived. Figure \ref{fig:exampleSigmoid} shows the $C(h_0)$ curve from the 339-441 Hz CasA search-and-recovery Monte Carlos, as  a representative example of the results obtained with this procedure. The uncertainties introduced by this procedure are within $10\%$.

For each of the targets and frequency bands we determine the sensitivity depth ${{\mathcal{D}}}^{90\%}$ \cite{GalacticCenterMethod, Dreissigacker:2018afk} of the search corresponding to $h_0^{90\%}(f)$: 
\begin{equation}
{{\mathcal{D}}}^{90\%}:={\sqrt{S_h(f)}\over {h_0^{90\%}(f) }}~~[ {1/\sqrt{\text{Hz}}} ],
\label{eq:sensDepth}
\end{equation}
where $\sqrt{S_h(f)}$ is the noise level associated with a signal of frequency $f$. For more details on how this is computed see Appendix \ref{A:PSD}. We find that the sensitivity depth values are $\approx$ constant in the low, medium and high frequency ranges, respectively, for each target; Table \ref{tab:sensDepth} shows these values. As the number of frequency and spindown templates increases, the search becomes less sensitive. For this reason at fixed frequency, the sensitivity depth is larger for the older targets (G347.3 is the eldest and then comes Vela Jr.) and for the same target, the sensitivity depth is higher at lower frequencies.

\begin{table}[t]
\begin{tabular}{|l|c|c|c|}
\hline
\hline
 \multicolumn{4}{|c|} {sensitivity depth $~{{\mathcal{D}}}^{90\%}~[{1/\sqrt{\text{Hz}}} ] $} \\
\hline
 & Cas A & Vela Jr. & G347.3 \\
\hline
\hline
 [20-250] ~~Hz & 60.3 & 76.5 & 82.9\\
 \hline
[250-520] ~Hz & 54.5 & 70.1 & 79.1\\
 \hline
[520-1500] Hz & 53.7 & 70.0 & 76.4 \\
 \hline
\hline
\end{tabular}
\caption{Sensitivity depth (\ref{eq:sensDepth}) corresponding to the $h_0^{90\%}$ upper limits set by this search. 
As the number of frequency and spindown templates increases, the search becomes less sensitive.
}
\label{tab:sensDepth}
\end{table}

We determine the 90\% upper limits with Eq. \ref{eq:sensDepth} using the measured ${S_h}$ and the values of ${{\mathcal{D}}}^{90\%}$ shown in table \ref{tab:sensDepth}. The results are shown in figure \ref{fig:ULs} and available in machine readable format in the Supplemental materials and at \cite{AEIULurl}. We note that the $h_0^{90\%}$ curves for the different targets (figure \ref{fig:ULs}), and all the curves derived from these (figures \ref{fig:epsilonULs}, \ref{fig:alphaULs}, \ref{fig:Ratio}), have the same shape. This is because we have used the sensitivity depth as scaling factor from the same measured ${S_h}$. Upper limits are not reported for frequencies that harbour spectral disturbances and were excluded from the search.

The most constraining upper limits are at $\approx$ 172.5 Hz for all targets and measure \lowestULc, \lowestULv~ and \lowestULg~ for CasA, Vela Jr. and G347.3, respectively.  These numbers are well in line with the predictions made in \cite{Ming:2017anf}.

The straight dashed line in figure \ref{fig:ULs} shows the so-called age-based spindown limit, which is the amplitude at each frequency that is consistent with a spin evolution due solely to gravitational wave emission, for the entire age of the object. We have taken $330$ yrs for Cas A, $5100$ yrs for Vela Jr. and $1600$ yrs for G347.3. Both the age and the distance of Vela Jr. are uncertain and typically two ``extreme'' scenarios are considered \cite{Ming:2015jla, Ming:2017anf, Abbott:2018qee, Allen:2014yra}, a Vela Jr. that is young and close (700 yrs, 200 pc) and a Vela Jr. that is old and far (5100 yrs and 900 pc\footnote{We note that this old age and farthest distance is from \cite{Allen:2014yra}. }). The first scenario yields a higher maximum gravitational wave amplitude but requires a broader search to investigate the extensive range of spindown parameters. The second scenario yields a smaller maximum signal amplitude (more than a factor of ten smaller), but the range of spindown values to search is smaller, which means that computational resources can be invested in depth rather than breadth. We have plotted the age-based upper limit for the older-farthest Vela Jr., because that is the most constraining. The search we carry out could detect a signal from the younger Vela Jr.

\subsection{Upper limits on the source ellipticity}

\begin{figure}[h!tbp]
   \includegraphics[width=\columnwidth]{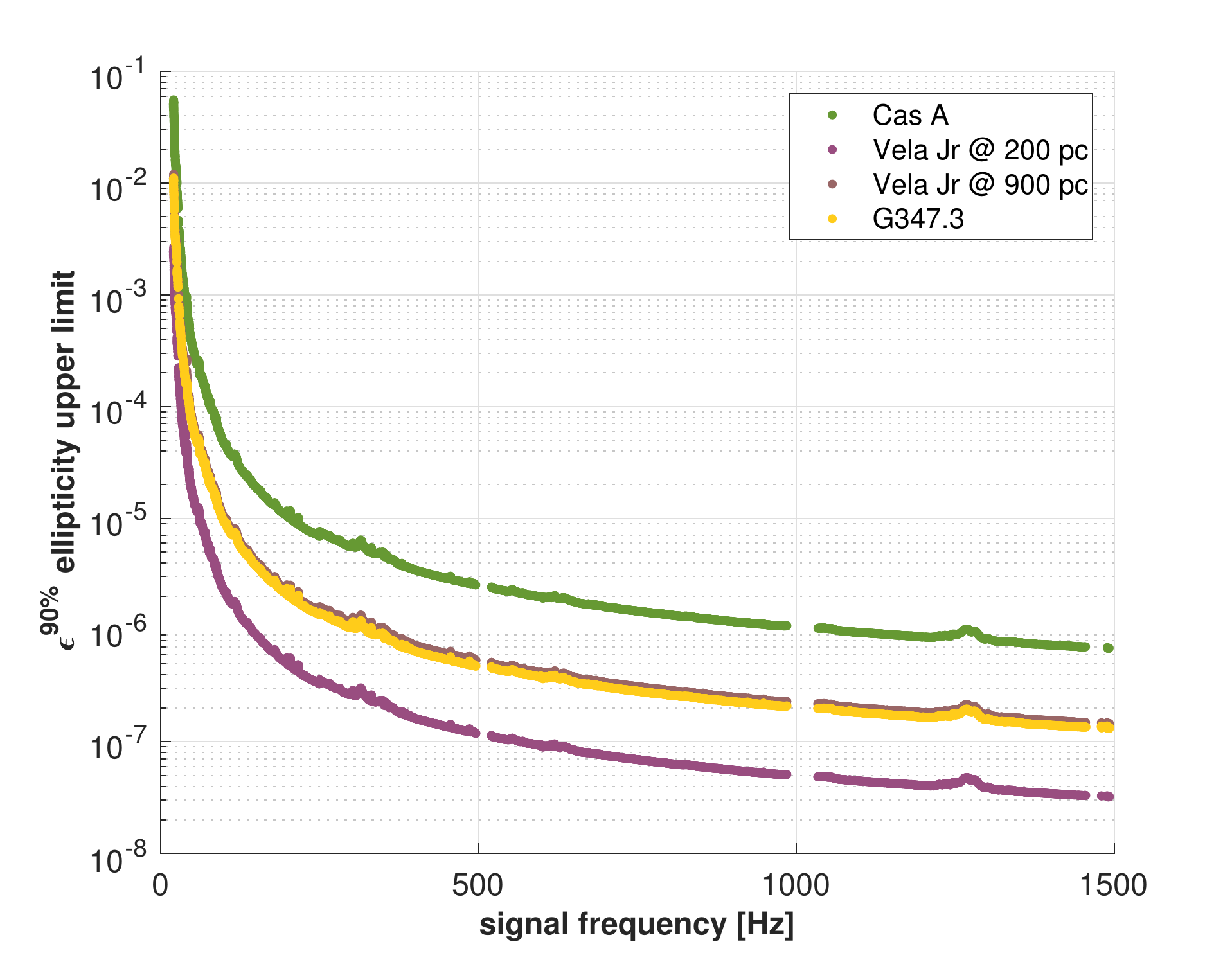}
\caption{ Upper limits on the ellipticity of the three targets. For Vela Jr. we show two curves, corresponding to two distance estimates: 200 pc and 900 pc.  For Cas A we assume 3.5 kpc and for G347.3, 900 pc.}
\label{fig:epsilonULs}
\end{figure}

The equatorial ellipticity $\varepsilon$ necessary to support continuous gravitational emission with amplitude $h_0$ at a distance $D$ from the source and at frequency $f$,  is 
\begin{equation}
\varepsilon = {{c^4}\over {4\pi^2 G}}{{h_0 D}\over {I f^2}}
\label{eq:epsilon}
\end{equation}
where $c$ is the speed of light, $G$ is the gravitational constant and $I$ the principal moment of inertia of the star. Based on this last equation, we can translate the upper limits on the gravitational wave amplitude in upper limits on the ellipticity of the source. The results are shown in figure \ref{fig:epsilonULs} assuming a fiducial value of the principal moment of inertia of $10^{38} \textrm{kg m}^2$. The upper limits can be scaled to any assumption for $I$ using Eq. \ref{eq:epsilon}. We use the following distance estimates for our targets: for Cas A, 3.5 kpc, for Vela Jr., 200 and 900 pc, and for G347.3, 900 pc. 

\subsection{Upper limits on the r-mode amplitude}

\begin{figure}[h!tbp]
   \includegraphics[width=\columnwidth]{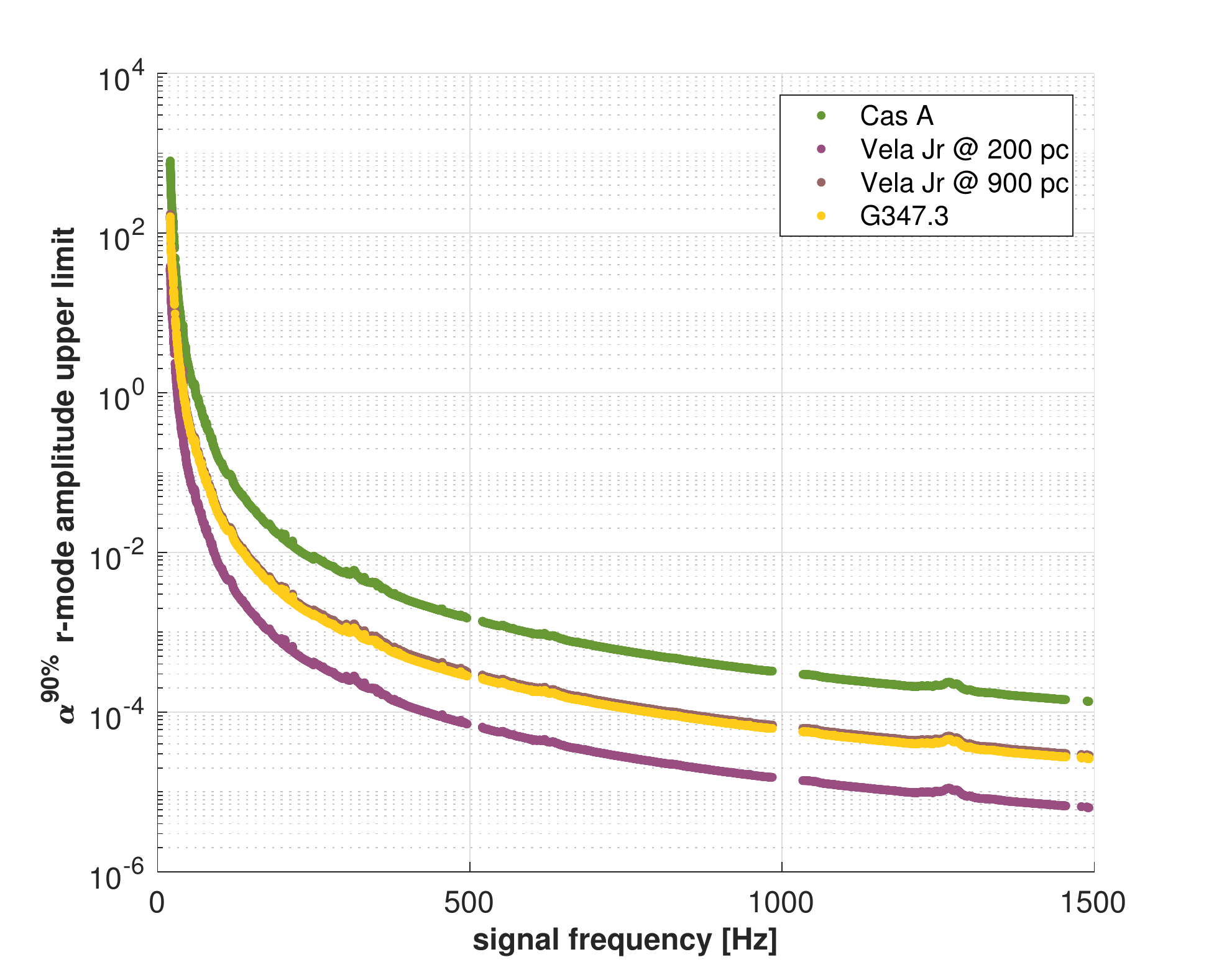}
\caption{Signal strain amplitude upper limits translated in upper limits on the r-mode amplitude.}  
\label{fig:alphaULs}
\end{figure}

Under a standard set of assumptions on the compact object, the r-mode amplitude $\alpha$ that would support continuous gravitational wave emission with amplitude $h_0$ at a frequency $f$, from a source at a distance $D$, can be written as \cite{Owen:2010ng}:
\begin{equation}
\alpha = 0.028 \left( {h_0\over{10^{-24}}}\right )\left ( {D\over{1~\textrm{kpc}}}\right ) \left ({{\textrm{100~Hz}}\over f} \right )^3 
\label{eq:rmodes}
\end{equation}
Using this relation we translate the amplitude upper limits in upper limits on the r-mode amplitude, as shown in figure \ref{fig:alphaULs}.

In the optimisation scheme used to determine the targets and the search set-ups (Eq. 17 of \cite{Ming:2017anf} and Eq.~\ref{eq:Priors} above) we assumed the largest  value of searched $\fddot$ to be consistent with a braking index of $n=5$. We note that this is a smaller value (by $\leq 30\%$) than could arise from r-mode emission at the spindown limit. On the other hand the  $\fddot$ range that we have used is independent of $\fdot$, and is taken to be equal to this maximum for all spindowns at a certain frequency.

\section{Conclusions}
\label{sec:conclusions}

\begin{figure}[h!tbp]
   \includegraphics[width=\columnwidth]{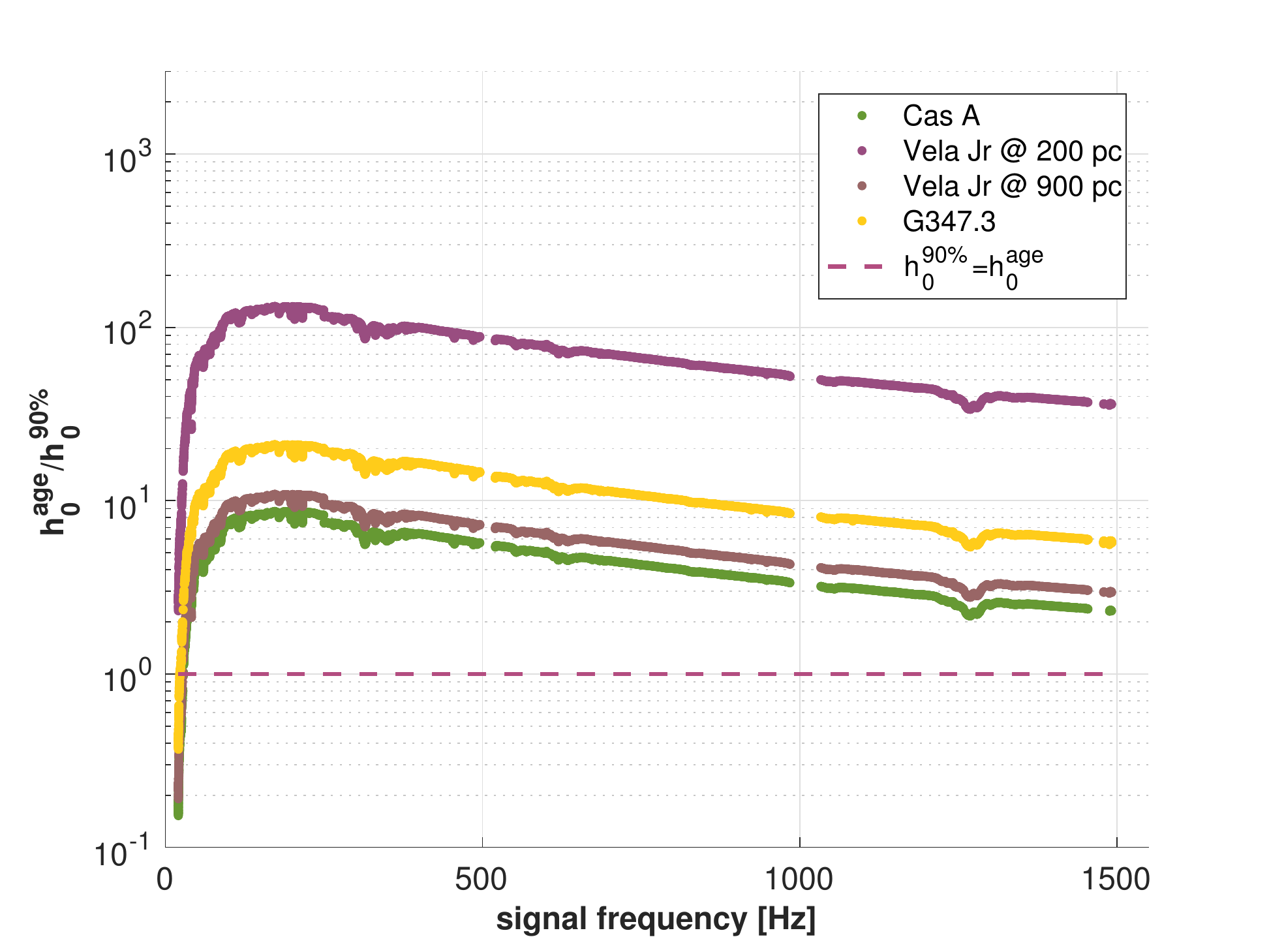}
\caption{Ratio of the age-based upper limit to the upper limits from this search. The larger this ratio, the more informative the result is.}
\label{fig:Ratio}
\end{figure}

We present the most sensitive results to date for continuous gravitational waves from Cas A, Vela Jr., and G347.3 We can exclude gravitational wave amplitudes as small as $9.5 \times 10^{-26}$ (G347.3) and ellipticities as small as $3\times 10^{-8}$ (Vela Jr. @ 200 pc). Above 28 Hz, our upper limits are below the indirect age-based amplitude upper limits for all three targets, and for G347.3 our results are below the indirect upper limit by a factor of 30 over more than 1400 Hz, as shown in figure \ref{fig:Ratio}. For frequencies higher than 200 Hz we can exclude that any of the objects have an ellipticity larger than $10^{-5}$, and Vela Jr., even at 900 pc, does not have an ellipticity larger than $10^{-6} (10^{-7})$ supporting gravitational wave emission at frequencies higher than $\sim$ 340 (1000) Hz. These constraints fall in a range of values that are plausible for compact stars, and significantly enhance the body of evidence that, if indeed there is a compact object at any of the locations that we targeted, its quadrupole moment {\it{is}} remarkably small.

This search is also the first concrete application of the optimisation procedure described in \cite{Ming:2015jla, Ming:2017anf} based on which the three targets Vela Jr., Cas A and G347.3 were chosen, and the associated search set-up. The procedure maximises the overall detection probability at fixed computing budget for the search. Priors on signal frequency, spindown, source ellipticity, age and distance need to be spelled out, and this is the difficult part. However, the interplay between these priors, the sensitivity of a search to the resulting distribution of signals, the detection probability and the computing cost is hard to factor-in based on simple qualitative arguments. 

One of the results that emerged from our analyses \cite{Ming:2015jla,Ming:2017anf} is that the detection probability is maximized by concentrating the computing power on the most promising three targets, Vela Jr., Cas A and G347.3.  This is not obvious, in fact \cite{Abbott:2018qee} not applying any optimisation scheme, search for emission from 15 supernova remnants, which means distributing the computing budget among more searches. Taking such an approach not only results in a different set of targets but also in search set-ups that are very different in comparison to ours, for the three common targets: for Cas A we perform 12 coherent searches, each $\approx$ 10 days long; \cite{Abbott:2018qee} do one search with a time baseline of $9$ days. For Vela Jr. we combine results from 8 searches, each having a coherent time baseline of  $15.4$ days; \cite{Abbott:2018qee} do one search and their longest coherent length is $9.3$ days. We search G347.3 by combining the results of 6 coherent searches, each over $20.4$ days, whereas \cite{Abbott:2018qee} do a single search over $6.7$ days.

\section{Acknowledgments}

The authors gratefully acknowledge the support of the many thousands of Einstein@Home volunteers, without whom this search would not have been possible. 

We thank Sin\'ead Walsh (U. Wisconsin Milwaukee) for useful insights on GCT search set-ups before mid August 2018; Divya Singh (Pennsylvania State U.) for sharing her expertise on the DM-off veto and Weichangfeng Guo (Beijing Normal U.) for the useful discussions and help in the search-and-recovery Monte Carlos.

M.A. Papa and B. Allen gratefully acknowledge support from NSF PHY grant 1104902. 

The authors thank the LIGO Scientific Collaboration for access to the data and gratefully acknowledge the support of the United States National Science Foundation (NSF) for the construction and operation of the LIGO Laboratory and Advanced LIGO as well as the Science and Technology Facilities Council (STFC) of the United Kingdom, and the Max-Planck-Society (MPS) for support of the construction of Advanced LIGO. Additional support for Advanced LIGO was provided by the Australian Research Council.
This research has made use of data, software and/or web tools obtained from the LIGO Open Science Center (\url{https://losc.ligo.org}), a service of LIGO Laboratory, the LIGO Scientific Collaboration and the Virgo Collaboration, to which the authors have also contributed. LIGO is funded by the U.S. National Science Foundation. Virgo is funded by the French Centre National de Recherche Scientifique (CNRS), the Italian Istituto Nazionale della Fisica Nucleare (INFN) and the Dutch Nikhef, with contributions by Polish and Hungarian institutes.

\newpage


\appendix
\newpage

\section{Clustering parameters}
\label{sec:clusteringParams}

The clustering procedure identifies multiple candidates as due to the same root cause; a single template is associated to each cluster and the follow-up stages act only on that candidate. This saves computational cost because only a single follow-up occurs for every cluster, rather than many separate follow-ups for each candidate within each cluster. 

We use the adaptive clustering procedure described in \cite{Singh:2017kss}. The clustering is tuned to a very low false dismissal for signals at the detection threshold, by means of fake-signal search-and-recovery Monte Carlos that mimic the actual search.  The resulting parameters are shown in Table \ref{tab:clustering}.

\begin{table}[t]
\begin{tabular}{|c|c|c|c|}
\hline
\hline
 & Cas A & Vela Jr. & G347.3 \\
\hline
\hline
\hline
 \multicolumn{4}{|c|} {20-250 Hz} \\
 \hline
 $~N_{F^*}~$ range &  \ClustNrangeCasAlow & \ClustNrangeVelalow  & \ClustNrangeGlow \\
 \hline
 $~C_{F^*}~$  &  \ClustCCasAlow & \ClustCVelalow  & \ClustCGlow \\ 
 \hline
$~P_{th^*}~$  &  \ClustPCasAlow & \ClustPVelalow  & \ClustPGlow \\ 
\hline
$~D_{th^*}~$  &  \ClustDCasAlow & \ClustDVelalow  & \ClustDGlow \\ 
\hline
$~G_{th^*}~$  &  \ClustGCasAlow & \ClustGVelalow  & \ClustGGlow \\ 
\hline
\hline  
 \multicolumn{4}{|c|} {250-520 Hz} \\
 \hline
 $~N_{F^*}~$ range &  \ClustNrangeCasAmid & \ClustNrangeVelamid  & \ClustNrangeGmid \\
 \hline
 $~C_{F^*}~$  &  \ClustCCasAmid & \ClustCVelamid  & \ClustCGmid \\ 
 \hline
$~P_{th^*}~$  &  \ClustPCasAmid & \ClustPVelamid  & \ClustPGmid \\ 
\hline
$~D_{th^*}~$  &  \ClustDCasAmid & \ClustDVelamid  & \ClustDGmid \\ 
\hline
$~G_{th^*}~$  &  \ClustGCasAmid & \ClustGVelamid  & \ClustGGmid \\ 
\hline
\hline
 \multicolumn{4}{|c|} {520-1500 Hz} \\
 \hline
 $~N_{F^*}~$ range &  \ClustNrangeCasAhigh & \ClustNrangeVelahigh  & \ClustNrangeGhigh \\
 \hline
 $~C_{F^*}~$  &  \ClustCCasAhigh & \ClustCVelahigh  & \ClustCGhigh \\ 
 \hline
$~P_{th^*}~$  &  \ClustPCasAhigh & \ClustPVelahigh  & \ClustPGhigh \\ 
\hline
$~D_{th^*}~$  &  \ClustDCasAhigh & \ClustDVelahigh  & \ClustDGhigh \\ 
\hline
$~G_{th^*}~$  &  \ClustGCasAhigh & \ClustGVelahigh  & \ClustGGhigh \\ 
\hline
\hline
\end{tabular}
\caption{Clustering parameters.}
\label{tab:clustering}
\end{table}

\section{Estimate of the power spectral density for the upper limit calculations}
\label{A:PSD}

Following Eq. \ref{eq:sensDepth} we compute the upper limits from the power spectral density $S_h$ and the sensitivity depth $\mathcal{D}$. The power spectral density is estimated by taking, for each detector, the mean power spectral density over the frequency interval spanned by the signal frequency in that detector during the observation time covered by the search.

\begin{figure}[h!bp]
   \includegraphics[width=\columnwidth]{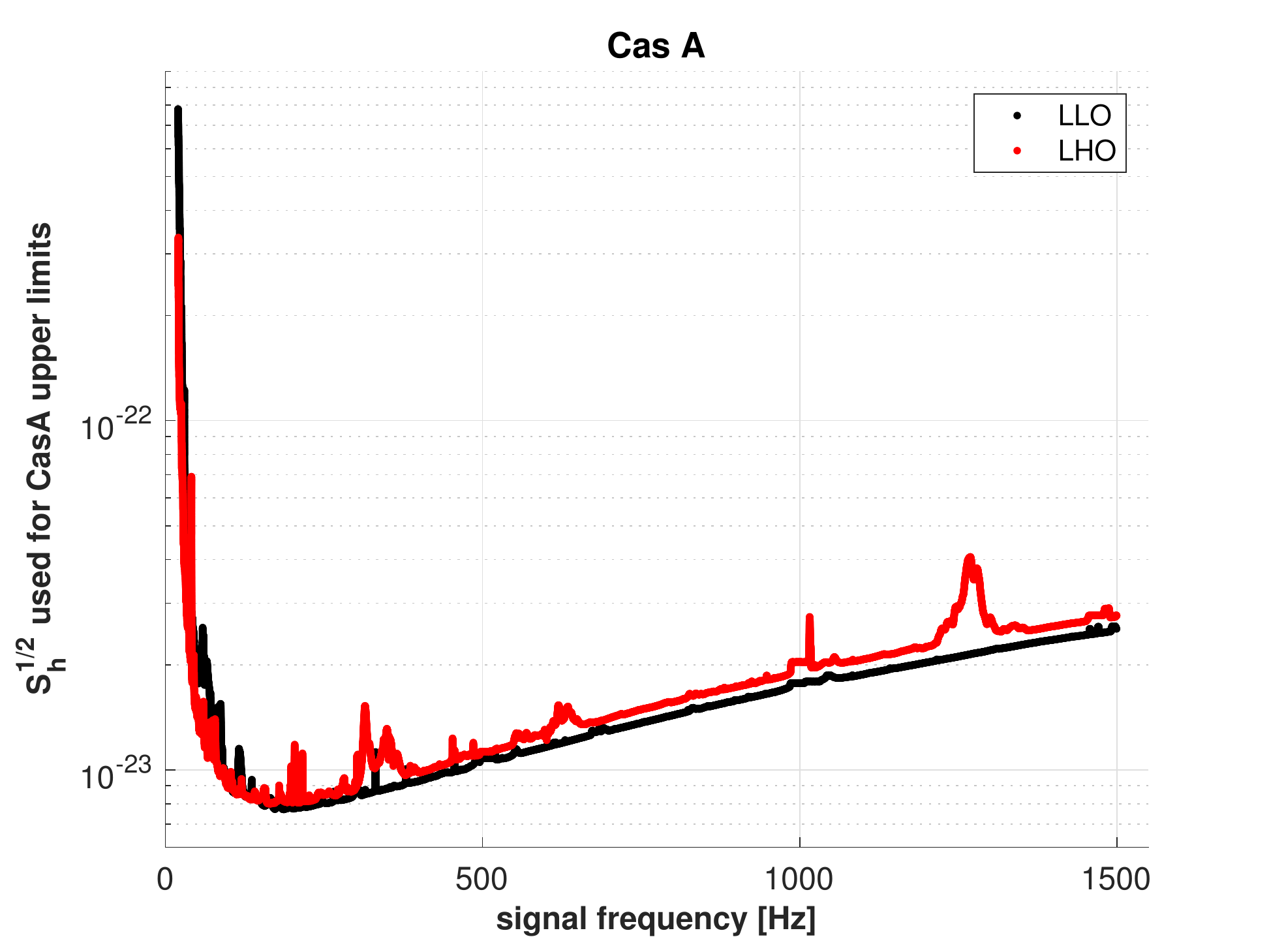}
\caption{$\sqrt{S_h}$ as function of signal frequency used in Eq.~\ref{eq:sensDepth} to compute the $h_0^{90\%}$ upper limits for Cas A.}
\label{fig:PSDs}
\end{figure}

When computing upper limits for signal frequency bands, typically half a Hz or more wide, the power spectral density is estimated over these frequency bands. In this case the difference between detector frequency and signal frequency can be overlooked because it is much smaller than the band over which the upper limit is valid and the upper limit is based on the loudest detection statistic value in that band. In this search, instead, we do not use the loudest candidate but the fixed stage-0 threshold, and set upper limits in signal frequency with the SFT resolution. In this case we need to identify which (SFT) frequencies have actually contributed to search for a signal at a given frequency for each target and each detector, and estimate the power spectral density in that range. Figure \ref{fig:PSDs} shows the power spectral density values {\it{as a function of signal frequency}} used to compute the upper limits for Cas A.

\end{document}